\documentclass[pra,aps,showpacs,twocolumn]{revtex4}
\usepackage{epsfig}
\usepackage{color}

\pacs{03.75.Gg,03.75.Kk}  

\begin{document} 
\title{Dephasing in coherently-split quasicondensates} 

\author{H.-P. Stimming$^1$,  N. J. Mauser$^1$, J. Schmiedmayer$^2$, I. E. Mazets$^{1,2,3}$}
\affiliation{$^1$Wolfgang Pauli Institute c/o University of Vienna, 1090 Vienna, Austria \\
$^2$Vienna Center for Quantum Science and Technology, Atominstitut, TU Wien, 1020 Vienna, Austria \\
$^3$Ioffe Physico-Technical Institute, 194021 St.Peterburg, Russia } 

\begin{abstract} 
We  numerically model the evolution of a pair of coherently split quasicondensates. 
A truly one-dimensional case is assumed, so that the loss of the (initially high) coherence between the two quasicondensates is due to dephasing only, but not due to the violation of integrability and subsequent thermalization (which are excluded from the present model). We confirm the subexponential time evolution of the coherence between two quasicondensates $\propto \exp [-(t/t_0)^{2/3}]$, experimentally observed by S.~Hofferberth {\em et. al.},  Nature {\bf 449}, 324  (2007). The characteristic time $t_0$ is found to scale as the square of the ratio of the linear density of a quasicondensate to its temperature, and we analyze the full distribution function of the interference contrast and the decay of the phase correlation. \\
\end{abstract} 
\maketitle 


\section{Introduction}

Dephasing and decoherence are phenomena at the heart of many-body physics which are 
deeply related to such fundamental 
problems as the crossover between quantum and classical dynamics of complex systems \cite{qcc}, 
reversibility of physical processes \cite{dirr} or 
quantum information storage and processing \cite{dqip}. To better understand dephasing and decoherence phenomena, 
we need systems, which are, on  one hand, simple and theoretically tractable, but, on the other hand, 
available experimentally. In particular, ultracold, weakly-interacting bosonic atoms offer such an opportunity. 
In the present paper, we investigate numerically  the time-dependent dephasing of 
ultracold atomic systems.

A system of identical bosons confined to one-dimensional (1D) geometry is experimentally 
realizable with ultracold atoms trapped on atom chips \cite{reviews} or in an array of tight 
waveguides formed by a 2D optical lattice \cite{dw1}. The conditions of 
one-dimensionality are smallness of the temperature with respect to the energy quantum $\hbar \omega _\mathrm{r}$ of the 
radial (harmonic) Hamiltonian, 
\begin{equation} 
k_\mathrm{B}T\ll \hbar \omega _\mathrm{r},    \label{xz.4} 
\end{equation} 
and smallness of the product of the 3D $s$-wave atomic scattering length $a_\mathrm{s}$ and the mean linear density $n_\mathrm{1D}$,
\begin{equation} 
n_\mathrm{1D} a_\mathrm{s}\ll 1.   \label{xz.5}
\end{equation} 
A 1D system of identical bosons interacting via contact pairwise potential is 
describable in the second quantization representation by the Hamiltonian 
\begin{equation} 
{\hat{\cal H}}= \int dz\, \left( \frac {\hbar ^2}{2m} \frac {\partial \hat{\psi }^\dag }{\partial z} 
\frac {\partial \hat{\psi } }{\partial z}+\frac {g_\mathrm{1D}}2 \hat{\psi }^\dag \hat{\psi }^\dag \hat{\psi }\hat{\psi }\right) , 
\label{xz.1} 
\end{equation} 
where $m$ is the mass of the boson, $\hat{\psi }=\hat{\psi } (z,t)$ is the bosonic annihilation field, and $g_\mathrm{1D}$ is 
the effective coupling strength of the 1D contact interaction (in what follows we assume repulsive interaction, 
$g_\mathrm{1D}>0$). The system described by this Hamiltonian supplemented by periodic boundary conditions is known to be 
fully integrable with the ground state properties and excitation spectrum in the thermodynamic limit given by the 
well-known Lieb-Liniger model \cite{LL}. 

Recalling that the coupling strength satisfies
$g_\mathrm{1D}=2\hbar \omega _\mathrm{r}a_\mathrm{s}$ as long as 
$a_\mathrm{s}\ll l_\mathrm{r}\equiv \sqrt{\hbar /(m\omega _\mathrm{r})}$ \cite{olsh}, 
we see that Eq. (\ref{xz.5}) requires smallness of the mean interaction energy per particle with respect to $\hbar \omega _\mathrm{r}$. 
Both Eqs. (\ref{xz.4}) and (\ref{xz.5}) mean small population of radially excited modes, either by temperature or interaction 
effects, respectively. 
 
In fact, the radial degrees of freedom are always excited {\em virtually}, with the excitation amplitude $\sim n_\mathrm{1D}a_\mathrm{s}$. 
This leads to the emergence of higher-orders in $\hat{\psi } ^\dag \hat{\psi }$ (cubic etc.) terms in Hamiltonian, in addition to 
what is given by Eq. (\ref{xz.1}). These terms, corresponding to many-particle (three-particle etc.) effective elastic collisions, 
violate the integrability and lead to thermalization on the time scale, at longest, 
$\sim 1/[\omega _\mathrm{r}(n_\mathrm{1D}a_\mathrm{s}^2/l_\mathrm{r})^2]$
\cite{recentwe}. Virtual radial mode excitations have been studied even earlier, in the context of 
soliton decay \cite{Muryshev} or quasi-1D (macroscopic) flow of a degenerate bosonic gas through a waveguide 
\cite{Salasnich1}. However, here we neglect this effect  in order to study purely integrable dynamics. 

In what follows, we consider uniform  ($n_\mathrm{1D}=\,$const for all $z$) 
and weakly interacting ($mg_\mathrm{1D}/\hbar ^2\ll n_\mathrm{1D}$) systems, ${n}_\mathrm{1D}=
\langle \hat{\psi } _j^\dag (z,t)\hat{\psi }_j (z,t)\rangle $ being the mean linear density of bosons. For temperatures 
below $T_\mathrm{qc}\sim (g_\mathrm{1D}\hbar ^2n_\mathrm{1D}^3/m)^{1/2}/k_\mathrm{B}$ \cite{qdcr}, 
a weakly-interacting system of bosons is in
the quasicondensate state \cite{Ha1}, which means that the operator $\hat{\psi }$ may be replaced by a classical 
complex-valued field with a phase fluctuating along $z$ (density fluctuations in the practically important 
long-wavelength range are suppressed via interactions).  {In this regime, not only the phase coherence is mantained, 
but also the density-density correlation function at zero distance approaches 1 (instead of 2, the value characteristic 
for a non-degenerate bosonic gas)} \cite{qdcr}.  
The stationary two-point 
correlation function of a quasicondensate at a finite temperature is \cite{Mora} 
\begin{equation} 
\left \langle \hat{\psi } _j^\dag (z,t)\hat{\psi }_j (z^\prime ,t)\right \rangle =n_\mathrm{1D} 
\exp \left( -|z-z^\prime |/\lambda _T \right) ,
\label{xz.2} 
\end{equation} 
where the thermal phase-correlation length is 
\begin{equation} 
\lambda _T =2\hbar ^2 n_\mathrm{1D} /(mk_\mathrm{B}T) .               \label{xz.3} 
\end{equation} 
Quantum noise is dominant on length scales shorter than $\lambda _T [mg_\mathrm{1D}/(\hbar ^2n_\mathrm{1D})]^{1/2}\ll \lambda _T$ 
\cite{Stimm1}. Finite-temperature fluctuations in degenerate bosonic gases in highly 
anisotropic traps has been extensively studied both 
theoretically \cite{P1,dri1} and experimentally \cite{han,dri2}.

One of the fundamental questions related to the Lieb-Liniger model is: how fast will two mutually decoupled quasicondensates 
with equal mean linear densities  ${n}_\mathrm{1D}$ decohere, if their initial fluctuations 
are highly correlated $\langle \hat{\psi } _1^\dag (z,0)\hat{\psi }_2 (z,0)\rangle \approx n_\mathrm{1D}$? 
Each of the quasicondensates 
has (upon tracing out the variables of another one) a finite temperature $T$.  
The measure of coherence will be the coherence factor 
\begin{equation} 
\Psi (t) =\langle \ \mathrm{Re}  \{ \ \exp [ i (
\varphi_1 (z,t)-\varphi_2 (z,t)-\theta_{\mathrm{ov}}(t)) ] \ \} \ \rangle,
\label{rz.1} 
\end{equation} 
where $\varphi_j (z,t)$ are the local phase operators for the condensate labeled $j$.
By $\theta_{\mathrm{ov}}(t)$ we denote  the constant overall phase of the system 
( {the phase associated with the Goldstone mode}) defined by
$\theta_{\mathrm{ov}}(t)=2\omega_r a_s \int_0^t dt' \int_0^L dz \left[ 
|\psi_2(z,t')|^4-|\psi_1(z,t')|^4 \right] $ which was described by Lewenstein and 
You  \cite{LevYou} (with $L$ the total length of the condensate). From now on, we replace 
operators $\hat{\psi }_1$, $\hat{\psi }_2$ by complex random functions ${\psi }_1$, ${\psi }_2$, 
whose fluctuations account for thermal noise. 

 {We use ``operational" definition (\ref{rz.1}) of the coherence factor on the two reasons: 
(1) density fluctuations are suppressed by interatomic repulsion and therefore Eq. (\ref{rz.1}) 
is practically equivalent to the more traditionbal definition} $\Psi (t) = n_\mathrm{1D}^{-1} 
\langle {\psi }_2^* (z,t) \psi _1 (z,t) e^{ -i\theta_{\mathrm{ov}}(t)}\rangle $  {(accounting for the 
correction to the overall phase); (2) in a time-of-flight interference experiment, the relative 
phase is directly measured, while the density fluctuations are much harder to detect.  
We retain the symbol of the real part in Eq. (\ref{rz.1}), because 
(i) the imaginary part of the expression 
in curly brackets becomes exactly zero only after averaging over an 
infinite ensemble of realization and is small but finite for real experimental data; 
(ii) we want to be consistent with the notation of Refs. \cite{BLD,EPJB9}.} 

 {The phase and density fluctuations in a quasicondensate are calculated from the harmonic approximation 
to the exact Hamiltonian (\ref{xz.1}). In the harmonic approximation, fluctuations are Gaussian with zero mean, 
and, hence, Eq. (\ref{rz.1}) reduces to} 
\begin{equation} 
\Psi (t) =\exp \left \{ -\frac 12 \langle [ 
\varphi_1 (z,t)-\varphi_2 (z,t)-\theta_{\mathrm{ov}}(t) ]^2 \rangle   \right \} .
\label{rz.2} 
\end{equation}

In experiment \cite{H1}, a subexponential decay of coherence 
\begin{equation} 
\Psi (t)\approx \exp [-(t/t_0)^\alpha ]   \label{xz.6} 
\end{equation}  
has been detected, with the numerical value of $\alpha $ statistically consistent with the hypothesis
\begin{equation} 
\alpha =2/3 .                    \label{xz.7} 
\end{equation}  
Initially there was a single quasicondensate of $^{87}$Rb atoms 
at the temperature $T_\mathrm{in}$ between 82~nK and 175~nK. Then it was coherently split into two quasicondensates 
with the density $n_\mathrm{1D}$ each (the values of $n_\mathrm{1D}$ were in the range from 20~$\mu \mathrm{m}^{-1}$ to  
52~$\mu \mathrm{m}^{-1}$). The splitting was made as adiabatic as possible, so that the initial fluctuation in both 
quasicondensates just after their full separation were almost identical. However, the coherence factor then decayed rapidly as the 
hold time grew, according to Eqs. (\ref{xz.6}, \ref{xz.7}), with $t_0\sim 0.01$~s, which is an order of magnitude shorter than 
the expected thermalization scale for a {\em non-degenerate} system of that density \cite{recentwe}. This obtained $t_0$ was in a 
fair agreement with the theory developed by Burkov, Lukin, and Demler \cite{BLD}, which will be 
briefly described later, under the assumption that the temperature 
of the two quasicondensates after the splitting was approximately equal to $T_\mathrm{in}$. However, the range of parameter 
variations (the radial trapping frequency was chosen to be either $2\pi \times 3.3$~kHz or $2\pi \times 4.0$~kHz) in Ref. 
\cite{H1} is too narrow to 
reliably determine the dependence of $t_0$ on $n_\mathrm{1D}$, $T$, and $\omega _\mathrm{r}$. 

\section{Theoretical approaches} 

The two existing theoretical descriptions  \cite{BLD,EPJB9} of dephasing in 1D quasicondensates share the common basic model 
but differ in technical tools to solve it. In both cases two quasicondensates, describable by Eq. (\ref{xz.1}), are separated 
by a potential barrier wide and high enough to make their tunnel coupling negligible and, hence, their time evolution  
after splitting fully distinct. These quasicondensates are 
assumed to have the same mean atomic density and placed in two waveguides with the same radial trapping frequency. 
The temperature is low enough ($k_\mathrm{B}T\lesssim \mu \equiv 
2\hbar \omega _\mathrm{r}n_\mathrm{1D}a_\mathrm{s}$) 
to consider only phononic part of the elementary excitation spectrum. 
 {Although static Eqs. (\ref{xz.2},~\ref{xz.3}) hold also for} $k_\mathrm{B}T> \mu $, 
 {the dynamic theories of Refs.  \cite{BLD,EPJB9} rely on the phononic type of the 
elementary excitations spectrum. In the case $k_\mathrm{B}T> \mu $ 
thermally populated particle-like modes contribute to the system dynamics, and we expect 
therefore a deviation of the low of coherence factor decay from the theoretical 
predictions} \cite{BLD,EPJB9}. 

The quantum noise is fully 
ignored, and excitations are represented by small-amplitude classical waves. Initially the fluctuations in the both 
quasicondensates are almost perfectly correlated (initial small interwell fluctuations serve as a seed noise, whose detailed 
properties are not ``remembered'' by the system in the long-time asymptotic regime and thus do not affect the final result).

The theory by Burkov, Lukin, and Demler \cite{BLD} was based on derivation of a Langevin-type equation with the random source term 
correlation and damping-term properties described by a certain kernel obtained by re-summation of diverging diagrams describing the 
exchange of quasiparticles between symmetric ($\hat{\psi }_+=(\hat{\psi }_1+\hat{\psi }_2)/\sqrt{2}$) and antisymmetric 
($\hat{\psi }_+=(\hat{\psi }_1-\hat{\psi }_2)/\sqrt{2}$) modes. 
Finally, Burkov, Lukin, and Demler obtained Eqs. (\ref{xz.6},~\ref{xz.7}) with $t_0=t_0^\mathrm{BLD}$, 
\begin{equation} 
t_0^\mathrm{BLD}=2.61\, \pi \hbar \mu {\cal K}/(k_\mathrm{B}T)^2 , 
\label{xz.8}
\end{equation} 
where ${\cal K}= \pi \sqrt{\hbar n_\mathrm{1D}/(2\omega _\mathrm{r}a_\mathrm{s})}$ is the Luttinger-liquid parameter of the system. 
Initially,  the symmetric mode is mostly populated, its initial temperature being $T_+\vert _{t=0}\approx 2T$, 
and only very small number of excitations 
are present in the antisymmetric mode, because of slightly nonadiabatic splitting. At large $t$, the temperatures of two modes 
equalize, $T_+\approx T_-\approx T$. Thus, in a strict sense, Ref. \cite{BLD} reproduces Eqs. (\ref{xz.6},~\ref{xz.7}) in the 
long-time asymptotic limit only, although in experiment this subexponential coherence decay law was observed even for $t<t_0$. 

The Burkov, Lukin, and Demler theory has, although, a weak point: it predicts overdamping of modes with energies larger than 
$\mu ^2/(k_\mathrm{B}T{\cal K})$. The damping rate of such modes is of the order of or larger than 
their frequency. Probably, 
this result is associated with a possible technical overestimation of the re-summed divergent 
series. 
It also may stem from the assumption of the purely linear dispersion law for
elementary excitations \cite{BLD} that provides a large phase space available
for products of a splitting of one phonon in the $+$ mode to two phonons in the
$-$ mode. However, higher order corrections to the linear phononic dispersion
law make this process energetically forbidden in 1D via a small but unavoidable
energy mismatch. Neglecting the latter fact may also results in obtaining too
fast dynamics of the system. This motivated us to reconsider the problem and to 
put forward an alternative explanation of the subexponential dephasing.

We refer a reader to our previous paper \cite{EPJB9} for the details of the calculations, which are briefly summarized below. 
We considered motion of pairs of compact wave packets of phonons (with the localization size of the order of the carrier wavelength
$2\pi /k$) in two random 1D media with relative fluctuations of local parameters (density and flow velocity) 
and follow their mutual dephasing, ascribing the obtained dephasing rate to the elementary mode with the momentum $\hbar k$. 
The source of mutual dephasing is the dependence of the phonon frequency $\omega _k$ on the local density $\delta n$ and flow velocity 
$\delta V$ fluctuations: 
\begin{equation} 
\omega _k = \left( c+\frac {dc}{dn_\mathrm{1D}}\delta n + \delta V\right) |k|,  
\label{xz.9}
\end{equation} 
where $c=\sqrt{\mu /m}\propto \sqrt{n_\mathrm{1D}}$ is the speed of sound. 
Then we assume that the contribution into the right-hand-side of Eq. (\ref{xz.9}) comes only from fluctuations with the wavelength 
longer than $2\pi /k$ (the influence of short-range fluctuations is averaged out). In other words, a propagating wave packet ``sees'' 
the fluctuations only on the length scales longer than its carrier wavelength. 

We calculate the statistical properties (the local correlation function at two different instants of time) 
of the {\em differences} of the fluctuations related to the 1st and 2nd 
quasicondensates, i.e., of $\delta n_1-\delta n_2$ and 
$\delta V_1-\delta V_2$ in the limit of asymptotically long time, when the 
symmetric and antisymmetric modes mostly equilibrate. The 
estimation for the dephasing rate $\Gamma _k$ 
for the two wave packets with the momentum $\hbar k$ propagating in two parallel quasicondensates gives 
\begin{eqnarray} 
\Gamma _k &\sim &c|k| \left[ \frac 58 \int _{-k}^k \frac {dk^\prime }{2\pi } \, \frac {k_\mathrm{B}T}{\mu n_\mathrm{1D}}\right] ^{1/2} 
\nonumber \\ &=& \varsigma \sqrt{\frac {k_\mathrm{B}T}{m n_\mathrm{1D}} } |k|^{3/2} . 
\label{xz.10}
\end{eqnarray}
Here we estimate $\varsigma \approx \sqrt{5/(8\pi )}\approx 0.446$. This estimation stems from our sharp-cutoff assumption. 
A different model, using some smooth function to eliminate the influence of modes with momenta $\gg \hbar k$, would give 
another value for $\varsigma $. However, later we shall see that $\varsigma \approx 0.446$ is quite a reasonable value. 
 
Note that Eq. (\ref{xz.10}) does not predict overdamping of the modes with the energies close to $k_\mathrm{B}T\lesssim \mu $: their 
damping rate is less than their frequency by a factor $\sim {\cal K}$.  

The shorter the wavelength of a mode, the faster this mode equilibrates. The integrated coherence factor is
then 
\begin{equation} 
\Psi (t) = \exp \left \{ -\frac {mk_\mathrm{B}T}{2\pi \hbar ^2n_\mathrm{1D}} \int _{-\infty }^\infty dk\, k^{-2} 
\left[ 1-\exp (-\Gamma _kt)\right] \right \} .
\label{xz.11} 
\end{equation} 
Since $\Gamma _k \propto |k|^{3/2} $ we obtain by integration $\Psi (t)= \exp [-(t/t_0)^{2/3}]$ but with 
$t_0=t_0^\mathrm{MS}$ \cite{EPJB9}, 
\begin{equation} 
t_0^\mathrm{MS} = \kappa \frac {\hbar ^3 n_\mathrm{1D}^2}{m(k_\mathrm{B}T)^2} =
\frac \kappa 4 \frac {m \lambda _T^2}\hbar , 
\label{xz.12} 
\end{equation} 
where $\kappa \approx 2.85 $ if we take $\varsigma =0.446$. 
The fact that $t_0^\mathrm{MS}$ does not depend on $c$ and, hence, on the atomic interaction strength in the limit ${\cal K}\gg 1$, 
seems to be deeply related to the independence of the thermal correlation length $\lambda _T$ [Eq. (\ref{xz.3})] on $c$ in the static regime.

\section{Numerical Model}

We directly simulated the time evolution of two $^{87}$Rb quasicondensates by solving (by a split step spectral method \cite{BaoMS}) two Gross-Pitaevskii equations with initial conditions chosen randomly corresponding to Bose-Einstein statistics of classical (thermal) excitations
in phase and density waves. The local phase $\phi _j$ and density $n_j$, $j=1,2$, values in the 1st and 2nd condensates are expressed through the respective values for the symmetric and antisymmetric modes (the local velocity is related to the phase as $V_j =(\hbar /m)\partial \phi _j/\partial z$). At $t=0$ they are   
\begin{eqnarray} 
\phi _{1,2}(z,0)&=& \frac {\phi _{+}(z,0)\pm \phi _{-}(z,0)}{\sqrt{2}}, \nonumber \\ 
n_{1,2}(z,0)&=&n_\mathrm{1D}+\frac {\delta n_{+}(z,0)\pm \delta n_{-}(z,0)}{\sqrt{2}}, 
\label{xza.1}
\end{eqnarray} 
where the fluctuations for the + and -- modes will be specified below. 

\subsection{Initial conditions: The splitting process}

Exact values of these fluctuations  depend on the details of the splitting process and its 
non-adiabaticity (for example, for the regime of the linear decrease of the tunnel coupling between two quasicondensates 
the population of phonons in the -- mode is expressible via Bessel functions \cite{gri}). However, without exact knowledge 
of the splitting process, we can only give an estimation of the initial fluctuations. In the present paper, we approximate the 
initial populations of the symmetric and antisymmetric elementary excitation modes by  thermal distributions with the 
temperatures $T_+$ and $T_-=\eta T_+$, $\eta \ll 1$, respectively. 
In the course of subsequent evolution,  the temperatures of the + and -- modes equalize and approach $[(1+\eta )/2]T_+$. 

The particular choice of this parameter ($\eta =0.1$ in our simulation) and the stability of our results against its 
variations are discussed in Sec.~IV.B. 

The initial fluctuations appearing in Eq. (\ref{xza.1}) are given by 
\begin{eqnarray} 
\delta n_\pm (z,0)& =&2\sqrt{\frac {n_\mathrm{1D}}L}\sum _k \sqrt{S_k}B_k^\pm \cos ({kz+\zeta _k^\pm }) , \nonumber \\
\phi _\pm (z,0)& =&\frac 1{\sqrt{{n_\mathrm{1D}L}}}\sum _k \frac 1{\sqrt{S_k}}B_k^\pm  \sin ({kz+\zeta _k^\pm }) , 
\label{xza.2} 
\end{eqnarray} 
where $S_k =|k|/\sqrt{ k^2 +4 m \mu / \hbar ^2 }$. 
Here the sum is taken over $k=2\pi M_k/l$, where $M_k$ is a non-zero integer number running from $-M_\mathrm{max}$ to 
$M_\mathrm{max}$. The phases $\zeta _k^\pm $ are uniformly distributed between 0 and $2\pi $ and $B_k^\pm $ is a positive number 
whose square has the exponential probability distribution $dP(|B_k^\pm |^2 ) d|B_k^\pm |^2 =\langle |B_k^\pm |^2 \rangle ^{-1} 
\exp \left( -|B_k^\pm |^2 / \langle |B_k^\pm |^2 \rangle \right) $. 

In the present paper, we completely neglect the quantum noise (zero-point oscillations of 
quantized local density and phase), since  taking it into account severely limits the time scale of reliable numerical integration of 
equations of motion of the system in the truncated Wigner approximation \cite{Sinatra}. Since we take into account only classical 
(thermal) excitations, we have: 
\begin{equation} 
\langle |B_k^\pm |^2 \rangle = \frac {k_\mathrm{B}T_\pm }{\sqrt{\frac {\hbar ^2k^2}{2m}\left( \frac {\hbar ^2k^2}{2m}+2\mu \right) }},  
\label{xza.3} 
\end{equation}
Here is the difference between our approach and that of Bistritzer and Altman \cite{AltB}, who simulated the dephasing of 
two quasicondensates with only quantum ($T_-=0$) fluctuations in the antisymmetric mode  at $t=0$. 
Using pseudo-random numbers $\xi _{k,1}^\pm $ and $\xi _{k,2}^\pm $, uniformly distributed between 0 and 1, 
we obtain the amplitude and phase of the excitation in a given mode: 
\begin{equation} 
B_k^\pm  = \sqrt{ \langle |B_k^\pm |^2 \rangle |\ln \xi _{k,1}^\pm |} , \quad \zeta _k^\pm =2\pi \xi _{k,2}^\pm . 
\label{xza.4}
\end{equation} 

\subsection{Numerical method}

For obtaining reliable random numbers for the choice of initial states, we employ a
Pseudo-Random Number generator of Wichmann and Hill \cite{wichm}, implemented here as a 
4-fold combined multiplicative congruential generator.
We use periodic boundary  conditions with a
simulation interval length which is of the order of the longitudinal extension  of the 
condensate. 

Then the time-dependent Gross-Pitaevskii equation is solved by a 4th-order Fourier Split-Step method, the corresponding time-dependent coherence factor is calculated and then averaged over a large enough sample of random realizations of this simulation. The resolution of the Fourier approximation is chosen such that the maximum energy of the quasiparticle (corresponding to the shortest resolvable wavelength) exceeds $k_\mathrm{B}T$, thus preventing the overestimation of the interwell coherence. 

\begin{figure}[t] 
\epsfig{file=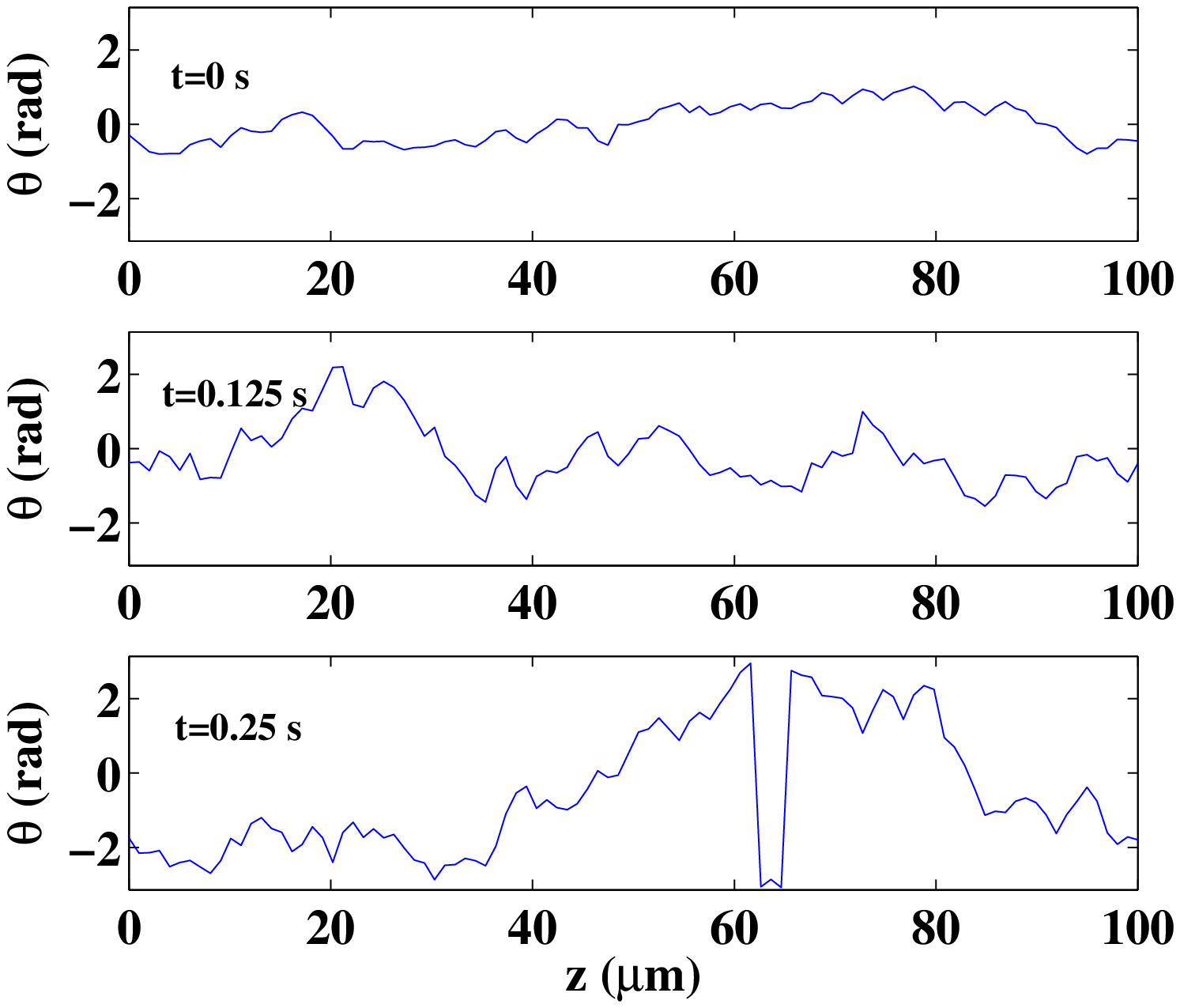,width=84mm}

\begin{caption} 
{\label{fig:typical_picture}
(Color online). Typical distribution of the relative phase $\theta(z,t)=\phi_1(z,t)-\phi_2(z,t)$
along the quasicondensate axis at several hold times between t=$0$ and $t=0.25$ s.
$n_{1D} = 50 \ \mu \mathrm{m}^{-1}$, $T_+ = 50$ nK.
}
\end{caption} 
\end{figure}

We obtained reliable (stable and convergent) results in the parameter range spanning over half an order of magnitude in 
both the temperature $T$ (from $30$ to $100$ nK) and the mean densities $n_\mathrm{1D}$ (from $30$ to $100$ $\mu$m$^{-1}$) with   
and additional constraint (to be discussed below): the reliable results were obtained for the density-to-temperature ratio 
less than or of order of 1~nK$^{-1}\, \mu \mathrm{m}^{-1}$, further increase of this ratio resulted in the appreciable 
dependence of the results on the grid size, which we could not eliminate within the reasonable range of the space and time steps. 
We also varied the effective 1D coupling strength by increasing the 
radial trapping frequency from$2\pi \times 3$ kHz to $ 2\pi \times 9 $  kHz.  Typical results are shown in Fig.  \ref{fig:typical_picture}.

\section{Numerical results}

\subsection{Full distribution function of interference}

\begin{figure}

\epsfig{file=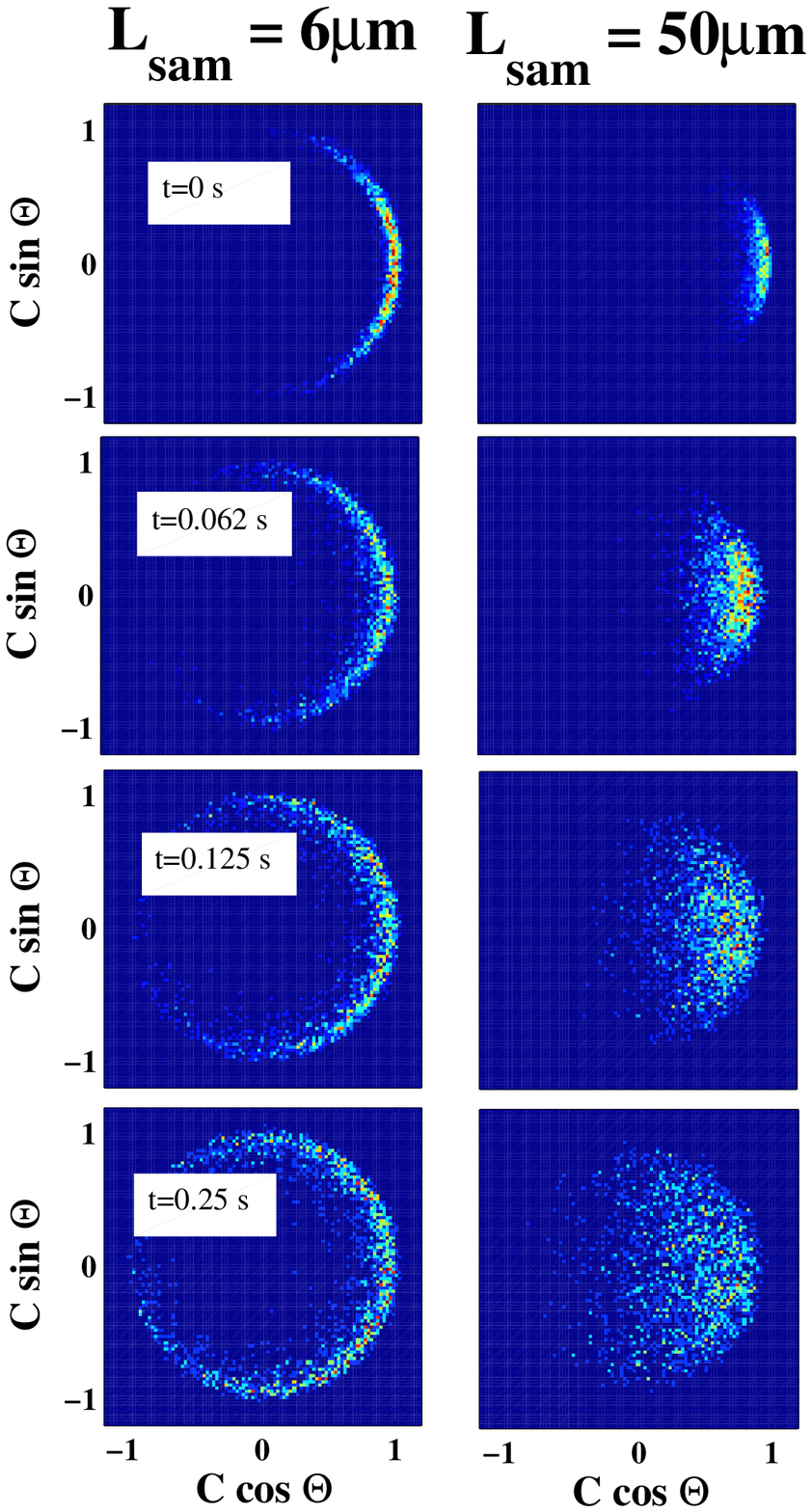,width=84mm}
\begin{caption} 
{\label{fig:8}
(Color online).
Time evolution of the Full Distribution Function shown as a false-color density plot 
(dark-blue: zero density; red: maximum density) derived from 3000 numerical runs. 
$n_{1D} = 50 \ \mu \mathrm{m}^{-1}$, $T_+ = 50$ nK.
Increasing time from top to bottom, times between 0 s  and 0.25 s. 
Sampling length:  $L_{\mathrm{sam}}=6\; \mu $m for left column,
$L_{\mathrm{sam}}=50\; \mu $m for right column.
Units on the axes are dimensionless. 
}   
\end{caption} 
\end{figure}

In Fig. \ref{fig:8}, we illustrate the numerically obtained temporal evolution of 
the  two-condensate joint 
Full Distribution Function (FDF), in the spirit of Ref.  \cite{Kita}. On polar plots shown in Fig. \ref{fig:8}, 
the polar radius of a point gives the contrast $C$ of the interference 
fringes \textit{integrated over the sampling length} $L_{\mathrm{sam}}$,  
and the polar angle is the phase $\Theta $  of such an integrated 
interference pattern. The FDF (in arbitrary units) is shown as a false color density plot. 

Locally, the relative phase randomizes rapidly, we can see that from Fig. \ref{fig:8}~(a). 
 {However, since the correlation length of the phase in each quasi-condensate is
$\lambda _T$, the contrast does not decrease significantly as long as 
$L_{\mathrm{sam}}\lesssim \lambda _T$. }  As we increase the sampling length, 
the local phases become more and more averaged out, and for the parameters of Fig. \ref{fig:8}~(b),
the contrast $C$ begins to decrease first, 
and then the distribution of $\Theta $ starts to spread. Remarkably, even at times as 
long as 0.25~s there are still many realizations yielding $C$ close to 1.

Note, that in our analysis we assumed   {equal mean atom numbers} in the two quasicondensate. 
The splitting process, that always happens in finite time, causes fluctuations of the relative number 
difference between two wells and, hence, provides an additional mechanism for the 
global phase diffusion \cite{JW97,LS98}.  {Experimentally observed \cite{H1} 
high initial phase coherence signifies uncertainty of the interwell atom-number difference 
(otherwise the overall phase would be completely random).} 
However, the phase diffusion affects only the global phase and not the coordinate-dependent noise 
and correlation properties. The global phase can be eliminated during the elaboration of 
experimental data and is therefore 
not  a major hurdle to experimental studies of dephasing. 


\subsection{Evaluation of the coherence factor }

\begin{figure}

\epsfig{file=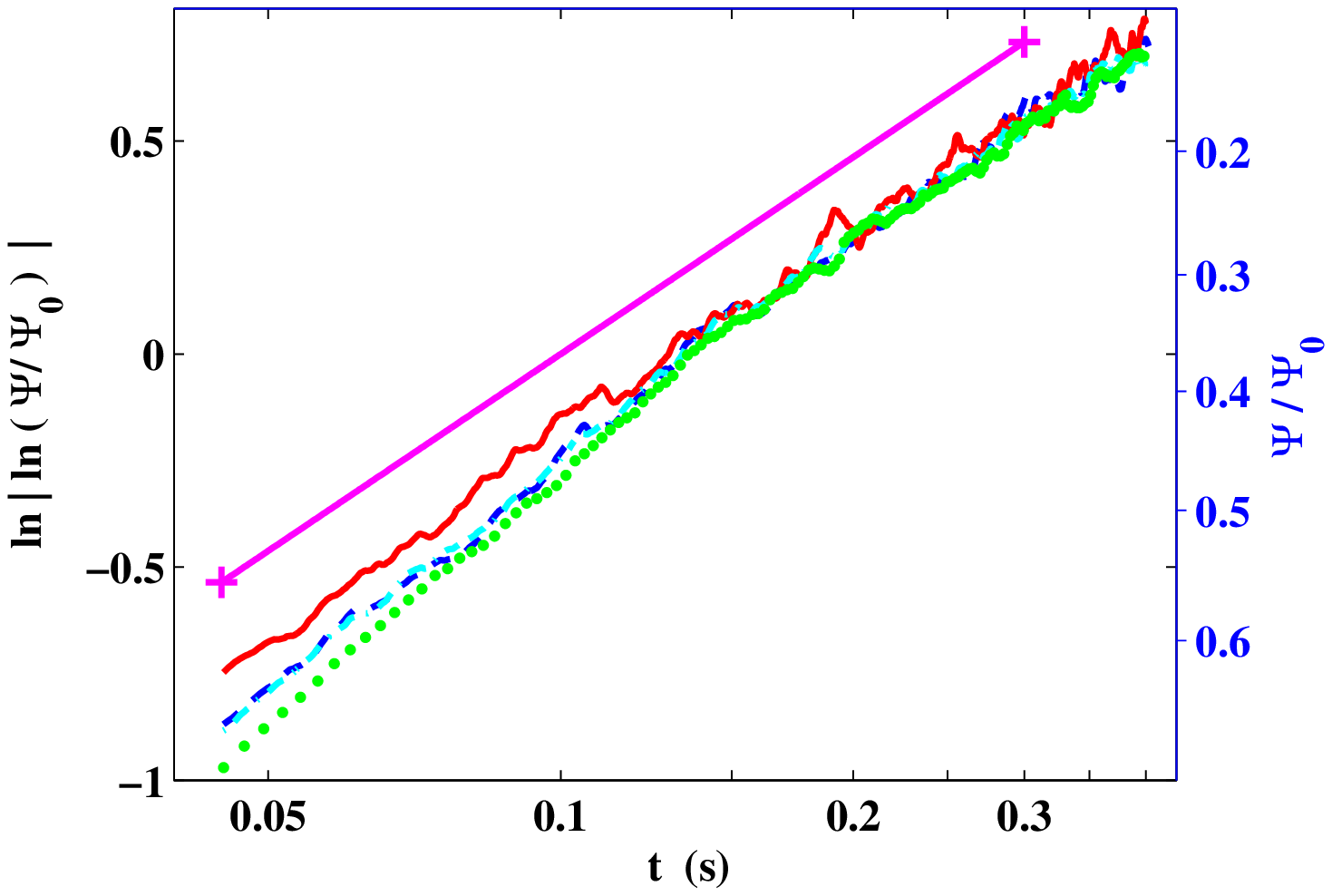,width=84mm}
  
\begin{caption}
{\label{fig:decay coherence factor}
(Color online). $\ln | \ln  \Psi|$
as a function of time for different values of $\eta = \frac{T_-}{T_+}$:
$\eta=0.14$ (red solid line), 0.1 (blue dashed line),
0.08 (cyan dash-dotted line), and 0.06 (green dotted line).
Logarithmic scale for time.
Straight solid line with crosses at the ends shows slope of $2/3$.
$n_\mathrm{1D}=50 ~\mu$m$^{-1}$ and $T_+=70 $~nK for all curves.}
\end{caption}

\end{figure}

Evaluating our simulated data, we obtained a subexponential decay of the coherence factor consistent 
with Eqs. (\ref{xz.6}, \ref{xz.7}) 
in a wide range of parameters,  {see Fig}.~\ref{fig:decay coherence factor}. 

 {Before proceeding further, we discuss our choice of the parameter $\eta $, 
the initial ratio of the temperatures in the $+$ and $-$ modes, and the sensitivity of 
our results to changes of this parameter. }

 {For all the  simulations, presented beginning from Fig. \ref{fig:2}, 
we used $\eta=0.1$. 
We choose this value because it gives the initial contrast that agrees with 
the experimental data \cite{H1} [$\Psi (0)\approx 0.9$ 
for a condensate of the length} $\sim 100~\mu $m].
 {Anyway, a  description of the initial fluctuations beyond our model 
of thermal fluctuations at the temperature $T_-=\eta T_+$ would require a detailed knowledge of the 
splitting process.} 
 
 {We checked the sensitivity of our simulations to the choice of the parameter $\eta$.
A plot of $\ln | \ln  \Psi |$ against $\ln t$ is shown in Fig. \ref{fig:decay coherence factor}. 
We can observe that the slope of the sub-exponential decay is varying only slightly. This means that 
the subexponential decay with $\alpha =2/3$ is quite insensitive to initial conditions within 
the given range of $\eta $. However, it remains 
unclear why this regime, which appears as asymptotic in both existing theories \cite{BLD,EPJB9}, sets on after a 
very short time (few} ms).

We now turn to the question of the decoherence time, and how it scales with the different parameters of the 1d system.
In Fig. \ref{fig:2}, we show numerically obtained values of $t_0$ as a function of $n_\mathrm{1D}$.
In Fig. \ref{fig:3}, we show  a plot of $t_0$ against the ratio $n_{1D}/T_+$. 
We discern two different ranges of parameters. If $n_\mathrm{1D}/T_+ \lesssim 1~\mathrm{nK}^{-1}\, \mu \mathrm{m}^{-1}$, 
we observed  a sub-exponential decay of $\Psi$ with the value of $\alpha$ from the interval
between 0.66 and 0.69. The corresponding decay time is fitted by the formula  
\begin{equation} 
t_0 \approx 6.4 \frac {\hbar ^3 n_\mathrm{1D}^2}{m(k_\mathrm{B}T_+)^2}= 1.6 \left. \frac {m\lambda _T^2}\hbar 
\right| _{T=T_+}. 
\label{xz.13}
\end{equation} 
Since in  our calculations we took $\eta = 0.1$, and, hence, the temperature of  both (+ and --) modes close 
to their equilibration is $T\approx 0.55\, T_+$, Eq. (\ref{xz.13}) corresponds to Eq. ({\ref{xz.12}) with $\kappa 
\approx 1.9$, which agrees by the order of magnitude with  $\kappa \approx 3$ 
 {approximately evaluated} in our theoretical model. 
Fig. \ref{fig:2}  shows  the fitting of $t_0$ by Eq. (\ref{xz.13}). 

\begin{figure} 

\epsfig{file=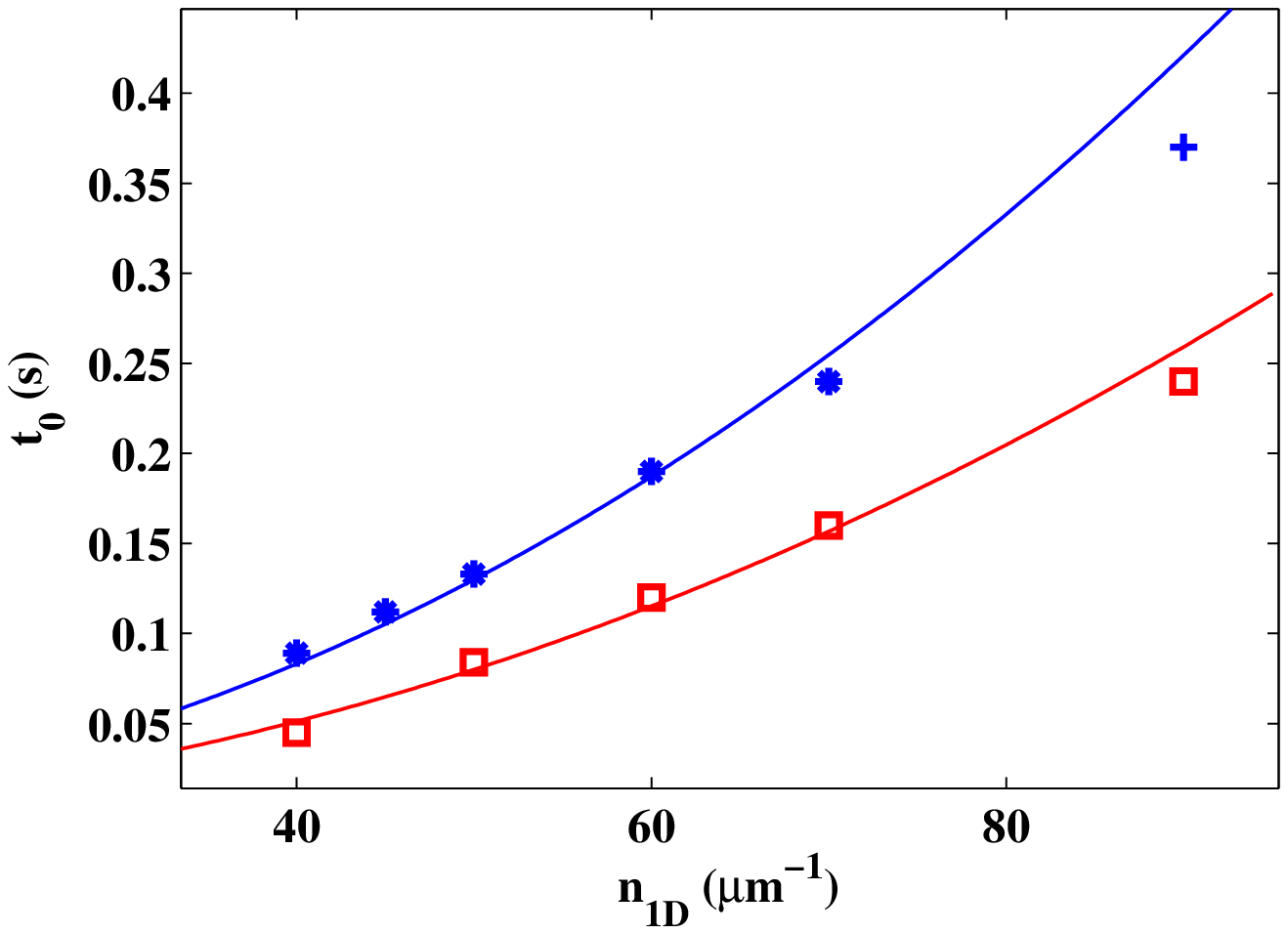,width=84mm}
\begin{caption} 
{\label{fig:2}
(Color online). Dephasing time $t_0$ 
as a function of mean density $n_\mathrm{1D}$ for different 
values of temperature $T_+$.
Symbols: simulation results, thin line: Eq. (\ref{xz.13}).
$T_+=70$ nK (blue asterisks and cross), and $90$ nK (red  squares).   }
\end{caption} 
\end{figure}

\begin{figure}[h] 
\epsfig{file=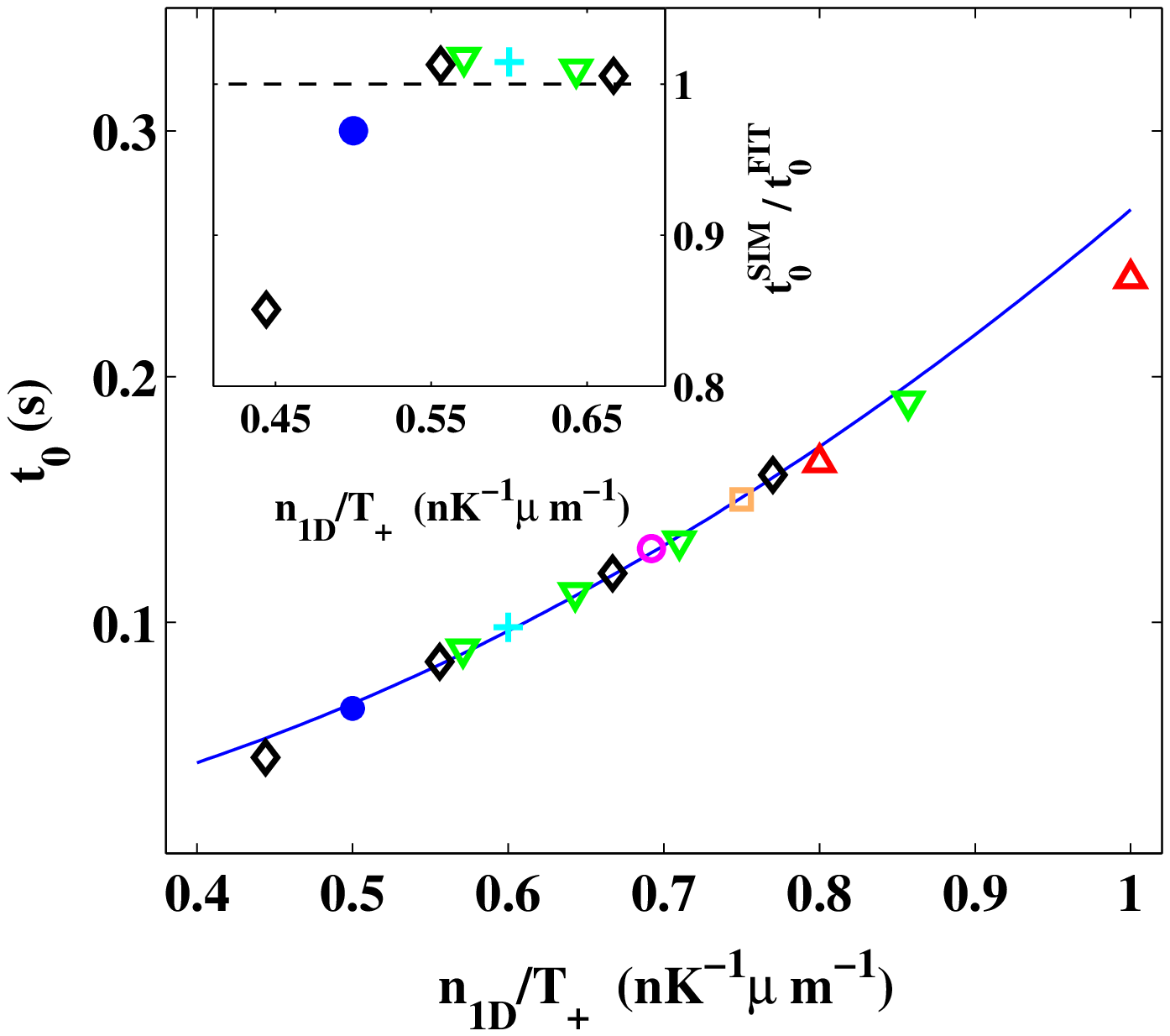,width=84mm}
\begin{caption} 
{\label{fig:3}
(Color online). Dephasing time $t_0$
as a function of the ratio $n_\mathrm{1D}/T_+$.
Solid blue line: Equation (\ref{xz.13}).
Symbols: simulation results, up triangles: $T_+=50$ nK, square: $T_+=60$ nK,
open circle: $T_+=65$ nK, down triangles: $T_+=70$ nK,
cross: $T_+=75$ nK, filled circle: $T_+=80$ nK,
diamonds: $T_+=90$ nK. Inset: the ratio of
numerically obtained values of $t_0$ to the values of $t_0$ given by the fitting Eq. (\ref{xz.13}).
The deviation of this ratio  by 15\%  from unity (horizontal dashed line) 
for the leftmost point is due to temperature high enough to significantly populate
particle-like states and thus violate the assumption of the phononic dispersion law
underlying Eq. (\ref{xz.13}). The next point shows the same tendency, but to less extent. }
\end{caption} 
\end{figure}
 
\begin{figure}[h] 
\epsfig{file=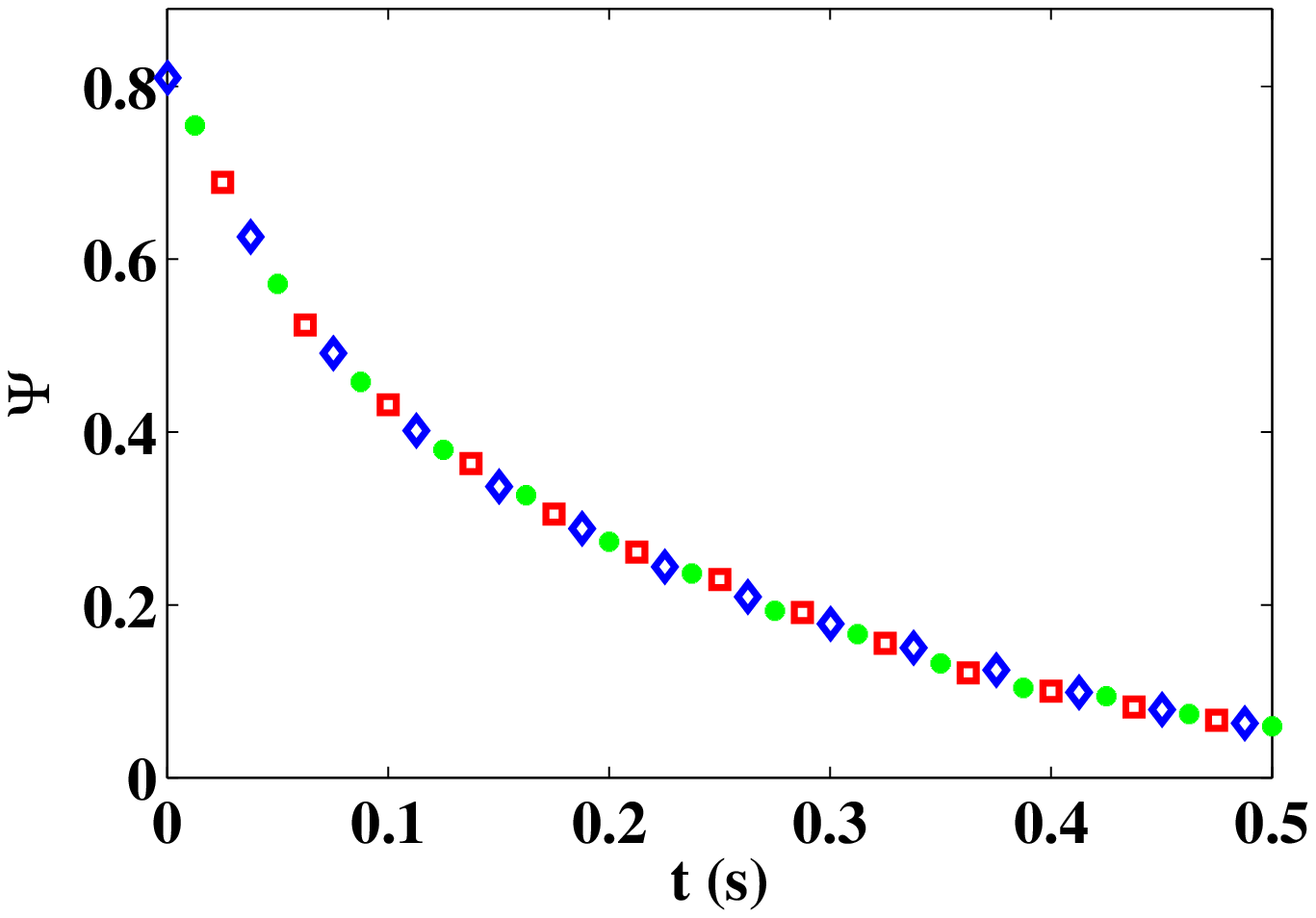,width=84mm}
\begin{caption} 
{\label{fig:6}
(Color online). $\Psi(t)$ for several values of
radial trapping frequency $\omega_r$,
$n_\mathrm{1D}=50 \ \mu \mathrm{m}^{-1}$ and $T_+=70$ nK. 
$\omega_r =  2\pi \times 3$ kHz (blue diamonds), 
$2\pi \times 6$ kHz (green asterisks), and   
$2\pi \times 9$ kHz (red squares). }
\end{caption} 
\end{figure}

For $n_\mathrm{1D}/T_+ \gtrsim 1~\mathrm{nK}^{-1}\, \mu \mathrm{m}^{-1}$, our numerical method starts to fail. 
We are still able to fit the contrast decay by the formula (\ref{xz.6}) with $\alpha \geq 0.7$, but the 
corresponding values of $t_0$ do not obey 
Eq. (\ref{xz.13})  anymore, and we have instead $t_0 \approx (m/\hbar ) \lambda _T^{1.67}b^{0.33}$,
where the length $b$ depends on the grid size, thus indicating a numerical artefact. A possible explanation of 
such a behavior is that for  small temperatures and high densities the interwell coherence persists for a long time, 
and the numerical error in simulations accumulates faster than the actual ``physical'' dephasing occurs, thus 
yielding the dephasing time dependence on the grid size. This explanation, yet not fully corroborated, is at least 
in accordance with the accelerated ($\alpha >0.7$) decay of simulated coherence in the problematic parameter range. 
The blue cross in Fig. \ref{fig:2} corresponds to a parameter set from this regime.

In the case where $k_\mathrm{B}T_+ > \mu $, $t_0$ starts to deviate from Eq. 
(\ref{xz.13}). The reason is that 
the theory resulting in Eq. (\ref{xz.12}) or (\ref{xz.13})  is based on the 
assumption that only the phononic part of the Bogoliubov spectra is occupied. 
This case corresponds to the first two leftmost values 
of Fig.  \ref{fig:3}.

In the range of applicability of Eq. (\ref{xz.13}),
we checked the independence of our result on the interaction strength, see Fig. 
\ref{fig:6} for a plot of $\Psi$ for different values of radial trapping. The only differences
visible in this comparison result from statistical fluctuations.
This independence of the time evolution of the coherence factor on the interaction strength 
confirmes the validity of the theoretical model proposed in Ref.  \cite{EPJB9}.

\subsection{Correlation function}

Additionally, we investigated correlation properties of the interwell coherence, which are an important tool of understanding the 
quasicondensate properties \cite{BT1D,H2,Stimm1}. Knowing the interwell coherence autocorrelation function one can, 
under Gaussian fluctuation assumption,  numerically construct distribution of the coherence factor (and the associated phase 
of the integrated interference pattern) for any sampling length. 
In contrast to these previous works, we deal now with non-stationary 
correlation properties. 
We express the  autocorrelation function of the interwell coherence via a new function $\Phi (z-z^\prime ,t)$ as 
\begin{equation} 
\langle \hat{\psi }^\dag _1 (z) \hat{\psi } _2 (z) \hat{\psi }^\dag _2 (z^\prime ) \hat{\psi } _1 (z^\prime )\rangle 
\equiv n_\mathrm{1D}^2 \exp (-\Phi ) , 
\label{xz.14} 
\end{equation} 
$\Phi $ being determined mainly by phase fluctuations (density fluctuations are 
suppressed by inter-atomic repulsion in 
the phononic regime).  {Analogously to Eq. (\ref{rz.2}), we may write }
\begin{equation} 
\Phi  = -\frac 12 \langle [ \varphi_1 (z,t)-\varphi_2 (z,t)-
\varphi_1 (z^\prime ,t)+\varphi_2 (z^\prime ,t) ]^2 \rangle .
\label{rz.3} 
\end{equation} 

 {We calculate the right-hand-side of Eq. (\ref{rz.3}) by generalizing the 
formula obtained by Whitlock and Bouchoule \cite{BT1D}, where we 
set the interwell tunnel coupling to zero and assume that, in the transient regime, 
mean population of each antisymmetric mode is given by its own temperature $T_k(t)$, which 
depends on the mode momentum $\hbar k$ and 
evolves in time approximately as} \cite{EPJB9} 
\begin{equation}   
T_k (t) =\frac {T_- -T_+}2\exp (-\Gamma _k t) +\frac {T_- +T_+}2  
\label{xz.15}
\end{equation} 
[cf. Eq. (\ref{xz.11}), where $\eta =\frac {T_-}{T_+} \ll 1$ is neglected]. 
Thus we write 
\begin{equation} 
\Phi =\frac 2\pi \int _0^\infty dk\, \frac {mk_\mathrm{B}T_k(t)}{\hbar ^2 n_\mathrm{1D}} 
(1-\cos k|z-z^\prime |) . 
\label{rz.4} 
\end{equation} 
 {We expand $\cos k|z-z^\prime |$ in series in powers of its argument, perform 
integration over $k$ for each term expressing the integrals via 
the gamma-function $\Gamma (s)=\int _0^\infty dx\, e^{-x}x^{s-1} $ and re-assemble the 
obtained terms into hypregeometric series }
\begin{eqnarray} 
\, _n F_m(b_1,\dots ,b_n;c_1, \dots ,c_m;x) =1 + \frac {b_1b_2 \, \dots \, b_n}{c_1c_2\, \dots c_m}\frac x{1!} &+& 
\nonumber \\
\frac {b_1(b_1+1)b_2(b_2+1) \, \dots \, b_n(b_n+1)}{c_1(c_1+1)c_2(c_2+1)\, \dots c_m(c_m+1)}\frac {x^2}{2!}+\dots 
\, .&& \nonumber  
\end{eqnarray}
Finally, we  arrive at the following expression:
\begin{equation} 
\Phi =\frac{2|z-z^\prime |}{\lambda _T} \left[ 1-\frac {4(1-\eta )}{3\pi (1+\eta )} \Xi (|z-z^\prime |/a) \right] , 
\label{xz.16} 
\end{equation} 
where $\lambda _T$ is to be evaluated at the equilibrated temperature $T\equiv T_\infty =(T_+ +T_-)/2$. 
The new auxiliary function $\Xi $ depends on $\tilde{z}= |z-z^\prime |/a$ only, where 
\begin{equation} 
a=\left( \frac {k_\mathrm{B}T_\infty  \varsigma ^2t^2}{mn_\mathrm{1D}}\right) ^{1/3}.
\label{xz.17} 
\end{equation}  
$\Xi $  {can be expressed via hypergeometrc series as follows}: 
\begin{eqnarray} 
\Xi (\tilde{z})&=& \frac 1{\tilde{z}} \left \{ \frac { \Gamma ( {\frac{2}{3}}) \tilde{z}^2 }2  \,   
_3F_4 \left(  \frac {\mbox{\scriptsize 1}}{\mbox{\scriptsize 6}},  {\frac {\mbox{\scriptsize 5}}{\mbox{\scriptsize  12}}}, 
{\frac {\mbox{\scriptsize  11}}{\mbox{\scriptsize 12}}}; 
 {\frac {\mbox{\scriptsize 1}}{\mbox{\scriptsize 2}}},\frac {\mbox{\scriptsize 5}}{\mbox{\scriptsize 6}}, \frac {\mbox{\scriptsize 7}}{\mbox{\scriptsize 6}}, 
\frac {\mbox{\scriptsize 4}}{\mbox{\scriptsize 3}}; -\frac {\mbox {\scriptsize 4}}{\mbox{\scriptsize 729}} {\tilde{z}^6}\right) - \right. 
\nonumber \\ && 
\frac { \tilde{z}^4 }{24} \, _4F_5\left( \frac {\mbox{\scriptsize 1}}{\mbox{\scriptsize 2}}, 
\frac {\mbox{\scriptsize 3}}{\mbox{\scriptsize 4}},{\mbox{\small  1}}, 
\frac {\mbox{\scriptsize 5}}{\mbox{\scriptsize 4}}; \frac {\mbox{\scriptsize 5}}{\mbox{\scriptsize 6}}, 
\frac {\mbox{\scriptsize 7}}{\mbox{\scriptsize 6}}, \frac {\mbox{\scriptsize 4}}{\mbox{\scriptsize 3}},  
\frac {\mbox{\scriptsize 3}}{\mbox{\scriptsize 2}}, \frac {\mbox{\scriptsize 5}}{\mbox{\scriptsize 3}}; 
-\frac {\mbox {\scriptsize 4}}{\mbox{\scriptsize 729}} {\tilde{z}^6}\right) + \nonumber \\ && 
\Gamma \left( -\frac {\mbox{\scriptsize 2}}{\mbox{\scriptsize 3}}\right)   \times    ~\label{xz.18} \\  & & 
\left. \left[   1- \, _3F_4 \left( 
-\frac {\mbox{\scriptsize 1}}{\mbox{\scriptsize 6}},\frac {\mbox{\scriptsize 1}}{\mbox{\scriptsize 12}}, 
\frac {\mbox{\scriptsize 7}}{{\mbox{\scriptsize 12}}}; \frac {\mbox{\scriptsize 1}}{\mbox{\scriptsize 6}}, 
\frac {\mbox{\scriptsize 1}}{\mbox{\scriptsize 2}},\frac {\mbox{\scriptsize 2}}{\mbox{\scriptsize 3}},
\frac {\mbox{\scriptsize 5}}{\mbox{\scriptsize 6}}; 
-\frac {\mbox {\scriptsize 4}}{\mbox{\scriptsize 729}} {\tilde{z}^6}\right) \right] \right \} . 
\nonumber  
\end{eqnarray} 

$\Phi $ changes significantly on a time scale $t_0$. At $t=0$ we have $\Phi =[4\eta /(1+\eta )]\lambda _T^{-1}|z-z^\prime |\ll 
|z-z^\prime |/\lambda _T$. 
In the opposite limit $t\gg t_0$ we have $a\rightarrow \infty$, $\tilde{z}\rightarrow 0$ for any finite coordinate 
difference, $\Xi \rightarrow 0$, and therefore $\Phi = 2|z-z^\prime |/\lambda _T$. 

An important advantage of the autocorrelation function (\ref{xz.14}) is its independence of the global 
phase diffusion \cite{LevYou,JW97,LS98}. Therefore the theoretical predictions for the autocorrelation 
function (\ref{xz.14}) admit direct comparison to experiment, unlike the contrast $\Psi $, which needs 
a correction to the phase diffusion. If the latter effect can not be neglected, the direct comparison 
of experiment to theory in terms of the contrast $\Psi $ is hindered, because unambiguous reconstruction 
of the global phase shift from the measurements \textit{in each experimental run}  is practically 
very difficult. 

\begin{figure}[t] 

\epsfig{file=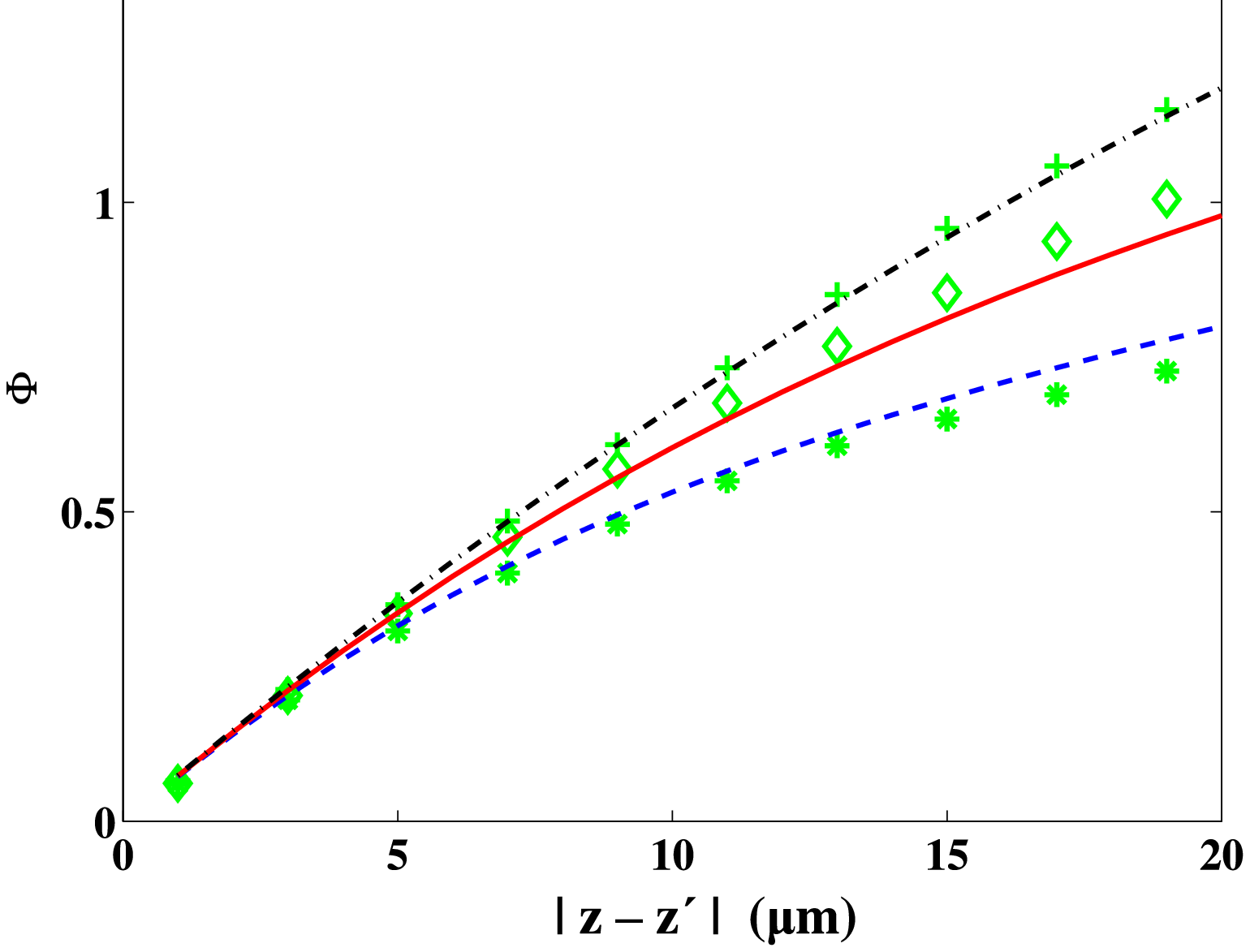,width=84mm}
\begin{caption} 
{\label{fig:7}
(Color online).
Function $\Phi$ (dimensionless) that defines the autocorrelation function of the interwell coherence Eq. (\ref{xz.14}) vs. the 
co-ordinate difference. Special symbols: numerical simulation results. Lines: fitting with $\varsigma =1.0$. 
The quasicondensate parameters are $n_\mathrm{1D}=40 \ \mu \mathrm{m}^{-1}$, $T_+ = 30$ nK. The time elapsed after 
the end of coherent splitting process is $t=0.1$~s (asterisks, dashed line), 0.2~s (diamonds, solid line), and 0.5~s 
(crosses, dot-dashed line). } 
\end{caption} 
\end{figure}

In Fig. \ref{fig:7} we display the simulated values of $\Phi $ and their fitting 
by Eqs. (\ref{xz.16}~-- \ref{xz.18}) at various times $t$. 
The only fitting parameter is $\varsigma $ whose value is found to be $\varsigma \approx 1.0$, which is in a fair agreement with 
the value $\varsigma \approx 0.74$ that corresponds to the numerical prefactor in Eq. (\ref{xz.13}) and is not too far 
from the  {estimated }value $\varsigma =0.446$.  

\subsection{Comparison to experiment} 

In the experiment \cite{H1} the temperature $T_\mathrm{in}$ 
{\em before} splitting was known, and the temperature $T_\mathrm{f}$ 
after splitting and equilibration between the + and -- modes was estimated 
using the theory of Ref. \cite{BLD}. Now we repeat the same procedure using our Eq. (\ref{xz.13}). 
Recalling that $\eta \ll 1$, 
we estimate $T_\mathrm{f}\approx 0.5\, T_+$, where $T_+$ is obtained from 
Eq. (\ref{xz.13}) where $t_0$ is given by the experiment for certain $T_\mathrm{in}$ and $n_\mathrm{1D}$ and 
present the results in Fig. \ref{fig:10}. Surprisingly, the values of $T_\mathrm{f}$ 
estimated from two models \cite{BLD} and \cite{EPJB9} 
are quite close to each other and are similar to $T_\mathrm{in}$. The  radial frequency range 
employed in \cite{H1} (from $2\pi \times 3.3$ kHz to 
$2\pi \times 4$ kHz) is too narrow to 
allow unambiguous determination of the dependence of $t_0$ on $\omega _\mathrm{r}$. 

\subsubsection*{Relation between $T_+$ and $T_\mathrm{in}$ under adiabatic splitting conditions} 
Obtaining the relation between $T_+$ and $T_\mathrm{in}$ is a subtle matter, not covered by the existing theories. 
Here we propose a way to estimate the ratio $T_+/T_\mathrm{in}$ from simple considerations.
The adiabaticity of the splitting process means that the 
entropy (but not the energy) is conserved. In other words, populations of the momentum states 
should not change in the course of splitting. Just after the splitting most 
of the noise occurs in the symmetric mode. However, the speed of sound in the + mode changes, as we show below. 

\begin{figure} 
\epsfig{file=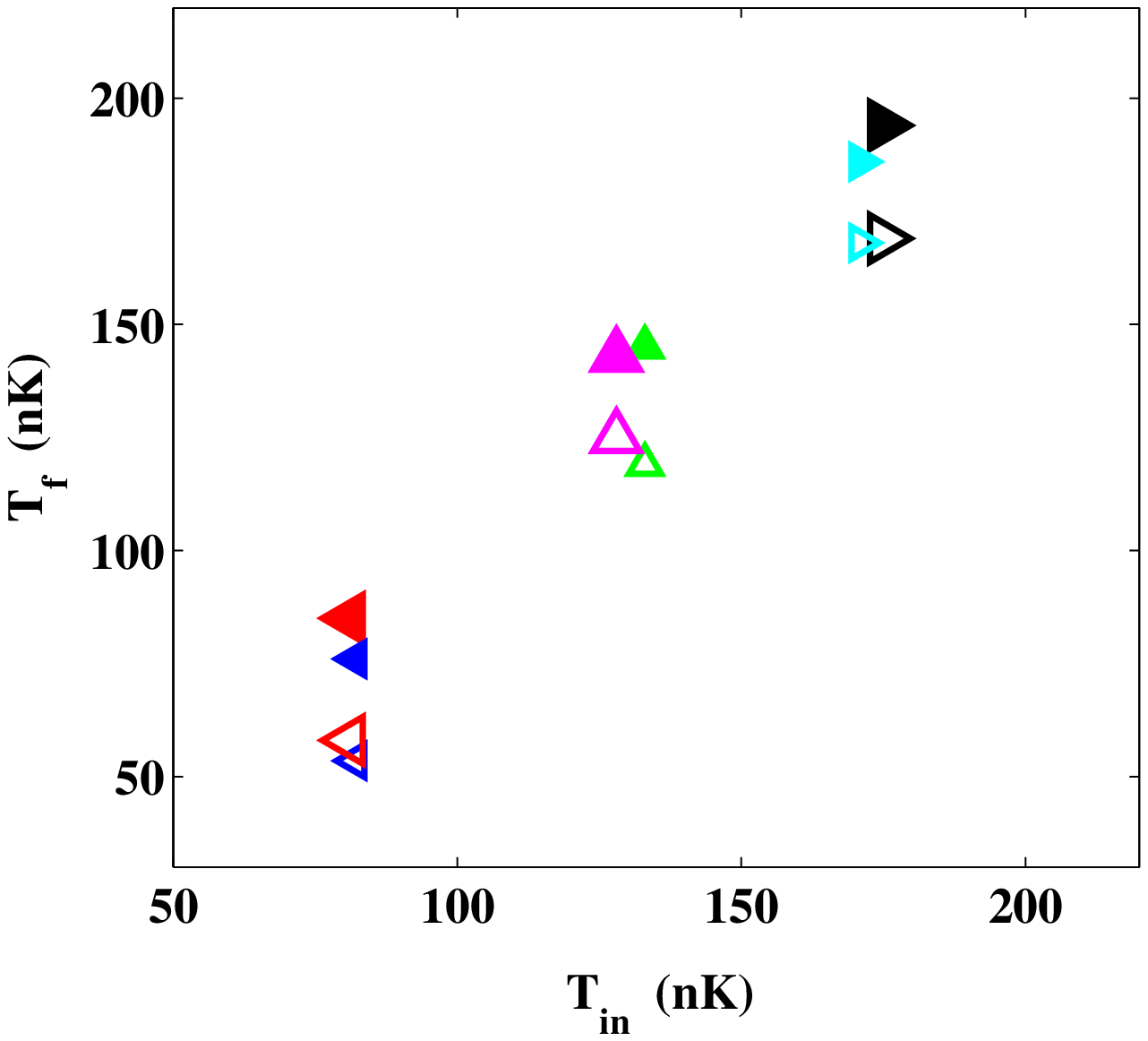,width=84mm}
\begin{caption} 
{\label{fig:10}
(Color online). Final temperature after equilibrating in experiment extrapolated by Eq.
(\ref{xz.13}) (open triangles) and by theory of Ref. \cite{BLD} (filled triangles). 
At radial frequency  $2\pi \times 3.3$ kHz (smaller symbols): background density
$n_\mathrm{1D}=20$ (left triangles),
$34$ (up triangles), and $52\ \mu \mathrm{m}^{-1}$  (right triangles).
At radial frequency  $2\pi \times 4.0$ kHz (larger symbols): background density
$n_\mathrm{1D}=22$ (left triangles),
$37$ (up triangles), and $ 51\ \mu \mathrm{m}^{-1}$ (right triangles).}
\end{caption} 
\end{figure}

Indeed, the speed of sound $c$ is to be calculated from the formula
\begin{equation} 
mc^2=\frac {4\pi \hbar ^2 a_s}m \int dx \int dy \, |\Psi _\perp (x,y)|^4 n_\mathrm{tot}, 
\label{xz.101} 
\end{equation}
where $\Psi _\perp (x,y)$ is the wave function (normalized to 1) 
of the ground state of transverse trapping Hamiltonian. To lowest order, we neglect the 
influence of the atomic interactions on the transverse profile of the quasicondensate \cite{Salasnich1}, in agreement 
with our assumption of a truly 1D regime. Eq. (\ref{xz.101}) applies for both single- and double-well potential shape, 
$n_\mathrm{tot}=2 n_\mathrm{1D}$ being the \textit{total} linear density of atoms in the system. 

The fundamental frequency of the transverse trapping potential is $\omega _\mathrm{r}$ before splitting. Therefore 
initially 
\begin{equation} 
\Psi _\perp ^\mathrm{in} (x,y)= (\sqrt{\pi } l_\mathrm{r})^{-1} \exp [-(x^2+y^2)/(2l_\mathrm{r}^2)] .  
\label{xz.102}   
\end{equation}
The splitting is designed so that in the end  the fundamental frequency of the 
radial oscillations in each of the two waveguides is again $\omega _\mathrm{r}$. Therefore,   
\begin{equation} 
\Psi _\perp ^\mathrm{as} (x,y)\approx [\Psi _\perp ^\mathrm{in} (x-w,y) +\Psi _\perp ^\mathrm{in} (x+w,y)]/\sqrt{2} 
\label{xz.103}   
\end{equation}
is to be substituted into Eq. (\ref{xz.101}) to determine the speed of sound after splitting. 
The separation $2w$ between the  two waveguides is so large 
that the overlap between the wave functions localized near $x=\pm w, \; y=0$ is negligible, as it must 
be in the zero-tunneling case. From Eqs. (\ref{xz.101}~-- \ref{xz.103}) one can easily see that the speed of sound 
drops after splitting by $\sqrt{2}$. 

As we discussed before, 
adiabatic splitting concerves the mode populations. If the temperature is comparable to the chemical potential, 
phonon-like modes (with the energy linearly proportional to the speed of sound) dominate. For each momentum 
$\hbar k$ the ratio $\hbar c |k| /T$ should be the same before and after the splitting 
(we denote the respective values of the speed of sound 
by ${c _\mathrm{in}}$ and ${c_\mathrm{as}}$). Therefore, we  expect 
\begin{equation}     
\frac {T_+}{T_\mathrm{in}}= \frac {c_\mathrm{as}}{c _\mathrm{in}}=\frac 1{ \sqrt{2}} .      
\label{xz.104}
\end{equation} 

However, $T_+$ can be affected by 
non-adiabatic effects, which are far beyond our 1D approach, such as heating via creation of 
vortices in the course of splitting and their 
subsequent decay \cite{heat} or heating by technical noise. Therefore it is 
extremely difficult to reliably predict \textit{a priori} 
the ratio $T_+/T_\mathrm{in}$ and, moreover, $T_\mathrm{f}/T_\mathrm{in}$.

\section{Conclusion}
 
To conclude, we simulated numerically the evolution of two coherently-split quasicondensates in a broad 
range of experimentally feasible densities, temperatures, and effective 1D coupling strengths 
(radial trapping frequencies). 
We reproduced the subexponential decay Eq. (\ref{xz.6}) of the interwell coherence with $\alpha $ 
very close to observed 
\cite{H1} and theoretically predicted \cite{BLD,EPJB9} value $2/3$. Our characteristic 
dephasing time $t_0$ varies quadratically 
with the ratio of the linear density to the temperature (as long as $k_\mathrm{B}T$ 
is smaller than the chemical potential and thus only the phonon part of 
the elementary excitation spectrum is thermally excited) and does not depend on the 
1D coupling strength. The latter fact 
has its static counterpart -- the independence of thermal phase-coherence length 
$\lambda _T$ of a quasicondensate on the speed of 
sound $c$ \cite{Mora}, see Eq. (\ref{xz.3}). In other words, the typical dephasing time 
scales as $t_0\sim m\lambda _T^2/\hbar $, in accordance with our model \cite{EPJB9}. This conclusion is corroborated by analysis 
of autocorrelation of the interwell coherence during the 
dephasing process. Although comparison to the experiment \cite{H1} does not show the clear preference of one of the two theories 
\cite{BLD,EPJB9} over another, our numerical simulations in a broad range of parameters unambiguously support our theory \cite{EPJB9}.

This work was supported by the the FWF (projects Z118-N16, P22590-N16, and SFB ``VICOM''), by the WWTF
(projects MA-45 and MA-07), and by the EC (STREP MIDAS).

\end{document}